\documentclass[letterpaper, 10 pt, conference]{ieeeconf} 

\IEEEoverridecommandlockouts                              
\usepackage{graphics} 
\usepackage{epsfig} 
\usepackage{mathptmx} 
\usepackage{amsmath} 
\usepackage{amssymb}  
\usepackage[table,xcdraw]{xcolor} 
\usepackage{array}
\usepackage{tabularx} 
\usepackage{booktabs} 
\usepackage{multirow} 
\usepackage{ragged2e} 
\usepackage{hhline} 
\usepackage{cellspace} 
\usepackage{titlesec} 
\usepackage{graphicx} 
\usepackage{enumitem} 
\usepackage{nicematrix} 
\usepackage{float} 
\usepackage{hyperref} 

\title{\LARGE \bf
Simple yet Effective Anti-windup Techniques for Amplitude and Rate Saturation: An Autonomous Underwater Vehicle Case Study}

\author{Pouria Sarhadi$^\dag$
\thanks{$^\dag$Pouria Sarhadi is with the School of Physics, Engineering and Computer Science, University of Hertfordshire, Hatfield, UK        ({\tt\small p.sarhadi@herts.ac.uk)}}}%

\definecolor{LR}{rgb}{1.0, 0.5, 0.5}  
\definecolor{DR}{rgb}{0.97, 0.0, 0.0}  
\definecolor{LG}{rgb}{0.5, 1.0, 0.5}  
\definecolor{DG}{rgb}{0.0, 0.8, 0.0}  
\definecolor{WH}{HTML}{FFFFFF}  
\definecolor{LB}{HTML}{E0F0FF}  
\definecolor{MB}{HTML}{C0D9FF}  
\definecolor{darkgreen}{rgb}{0.0, 0.5, 0.0}  

\begin{document}

\maketitle
\thispagestyle{empty}
\pagestyle{empty}

\begin{abstract}
Actuator amplitude and rate saturation (A\&RSat), together with their consequent windup problem, have long been recognised as challenges in control systems. Anti-windup (AW) solutions have been developed over the past decades, which can generally be categorised into two main groups: classical and modern anti-windup (CAW and MAW) approaches. Classical methods have provided simple and effective results, mainly addressing amplitude saturation. In contrast, modern approaches offer powerful and theoretically sound solutions capable of handling both amplitude and rate saturations. However, MAW's derivation process often imposes restrictive conditions and can be complex to apply in practical engineering problems. Nevertheless, the literature has paid limited attention (if not entirely ignored) to the potential of simple yet effective CAW schemes that can operate in the presence of both A\&RSat elements. This paper revisits this issue and proposes modifications to two well-known controllers: PID and LQI. The obtained results, benchmarked on the REMUS AUV yaw control problem and compared with constrained MPC, indicate that these classical techniques can still provide simple yet effective solutions with comparable performance, at least for SISO systems. These findings may stimulate further research into solutions that achieve comparable performance with only one (or a limited number of) additional tuning parameters and straightforward implementation.
\end{abstract}

\section{Introduction}
\vspace{-3pt}
Actuators are fundamental components of control systems. Input limitations, or saturation, are usually the first major nonlinearity encountered in practical systems, making even a linear control loop nonlinear. All physical actuation systems are subject to inherent saturation limitations. Challenges of saturating actuators have long been understood and discussed in the control engineering literature \cite{tustin1947Saturation,Lozier1956Saturation,bernstein1995chronological}. Traditionally, these issues are identified as \textit{integrator windup} in PI or PID (proportional-integral-derivative) controllers that include integral action \cite{fertik1967AWfirst,astrom1989AW,aastrom2021feedback}. Nevertheless, it was later understood that not only controllers with integral action may encounter windup, but also systems with relatively slow or unstable modes can exhibit windup behavior when subjected to actuator saturation \cite{doyle1987windup,kothare1994AW}. Thus, windup in a system can be interpreted as an inconsistency between the controller output and its internal states when the control signal saturates \cite{kothare1994AW}. 

Two types of constraints, amplitude and rate saturation, can occur in actuators. Neglecting the latter in control design has been shown to cause instability and, in some cases, catastrophic failures \cite{Dornheim1992Gripen,Shifrin1993Gripen,Stein2003Respect}. Gunter Stein's seminal work, \textit{Respect the Unstable} \cite{Stein2003Respect}, highlights the role of rate saturation in practical instability and identifies it as a principal cause of several tragic incidents, including the Chernobyl disaster and multiple aircraft crashes \cite{Stein2003Respect}. Among the most notable are the SAAB Gripen JAS-39, which suffered two accidents, and the YF-22 by Lockheed Martin \cite{Dornheim1992Gripen,Shifrin1993Gripen,Stein2003Respect,brieger2009flightATTAS}. These incidents motivated the research community to develop novel methods, leading to modern anti-windup solutions, with peak research activity during 2000–2020. MAW techniques based on optimal and robust control have been at the forefront of this effort, providing solid formal stability proofs \cite{tarbouriech2009ModernAWreview1,galeani2009ModenAWtutorialreview2,zaccarian2011AWBook1,tarbouriech2011AWBook2}. In the past decade, adaptive versions of MAWs have also been proposed \cite{sarhadi2016PIDAW1,sarhadi2016PIDAW2,sofrony2024AWadaptive}. However, these controllers impose specific design requirements, and many can only handle open-loop stable systems. Another issue, despite their theoretical appeal, is the need to solving Linear Matrix Inequalities (LMIs), which is not always attractive for engineers. This has driven increased interest in Model Predictive Control (MPC), given its ability to handle constraints. However, MPC implementation can be computationally expensive and demands a precise model.

Classical AW solutions, while demonstrating elegant and practical approaches that gained popularity in industry, have generally not addressed the issue of rate saturation. In recent years, there has been growing interest in ‘simple yet effective’ approaches that can be integrated with baseline controllers such as PID, enabling them to tackle practical control challenges \cite{skogestad2023advanced}. Nonetheless, classical AW solutions have seen comparatively less research activity \cite{visioli2003AWclassicapproaches,Visioli2006AWclassicapproaches}.  

The main contribution of this paper is revisiting the problem of rate saturation in CAW and proposing simple solutions that require only a single tuning parameter to handle R\&Sat, at least for single-input single-output (SISO) systems. This contrasts with almost all CAW solutions, which address only amplitude saturation, and MAW approaches, which can be complex to implement. The proposed techniques modify two popular controllers: the cascade PID (Proportional–Integral–Derivative) and the LQI (Linear–Quadratic–Integral). The obtained results show performance comparable to that of MPC, verifying their effectiveness. This can potentially open a new research avenue into (relatively) ignored classical AW solutions capable of operating effectively across a variety of systems, thereby assisting industrial engineers seeking simple controller structures.

The remainder of this paper is organised as follows. Section~\ref{section:problem_statement} briefly presents the problem with further elaboration. Section~\ref{section:problem_statement} introduces two AW modifications to CAW controllers capable of handling both A\&RSat. Section~\ref{section:simulation_results} presents the simulation results and a comparison between the modified techniques and MPC for the REMUS AUV yaw control problem. Finally, Section~\ref{section:conclusion} concludes the paper.
\section{CAW and MAW Solutions: Rate Saturation Problem}
\vspace{-3pt}
\label{section:problem_statement}
Reviewing all available AW solutions is beyond the scope of this paper; however, it is essential to provide some background to ease the introduction of the proposed techniques. Conventionally, saturation has been addressed by adopting conservative (low-gain) controllers, using only a portion of the actuator capacity in the nominal design. A common rule of thumb is to reserve approximately 50\% of the actuator capacity to accommodate uncertainties under perturbed conditions. Another conservative approach is to deliberately filter or slow down reference signals generated by guidance or planning algorithms in autonomous vehicles, which are also formalised under reference governor frameworks \cite{angeli2002referencegovernor}. Both strategies can reduce system responsiveness, which may be undesirable in high-performance autonomous systems. Therefore, a central goal in autonomous system control is to fully exploit actuator capabilities to maximise performance while maintaining robustness.

\subsection{Classical methods in dealing with saturation}
\vspace{-3pt}
Classical techniques for handling windup are usually \textit{ad hoc} solutions that impose limits on the integral operator or utilise feedback to manage saturation. In general, the problem is to control a system described by the following dynamics:
\begin{align}
\dot{x}_p(t) &= A_p x_p(t) + B_p u_c(t), \label{eq:state_space_1} \\
y(t) &= C_p x_p(t),
\label{eq:state_space_2}
\end{align}
where $x_p \in \mathbb{R}^{n}$ denotes the vector of system states, $\dot{x}_p$ its time derivative, and $y$ the system output. The subscript $p$ refers to the plant, distinguishing the open-loop system model. The matrix $A_p \in \mathbb{R}^{n \times n}$ defines the system dynamics and stability, while $B_p \in \mathbb{R}^{n \times 1}$ and $C_p \in \mathbb{R}^{1 \times n}$ are the input and output matrices, respectively. The control input $u_c$ generated by the controller is limited by amplitude and rate constraints, $|u_c| \le u_{\max}$ and $|\dot{u}_c| \le \dot{u}_{\max}$, and after passing through the actuator, $u_c$ becomes $u_{ac}$. It should be noted that a SISO system representation is considered. We present two popular categories of CAW here:
\subsubsection{\textit{Ad-hoc} methods: imposing limits on the integral operator:}
\label{section:ad-hoc-amplitude-AW}
\begin{figure}[t!]
	\centerline{\includegraphics[width=0.5\textwidth]{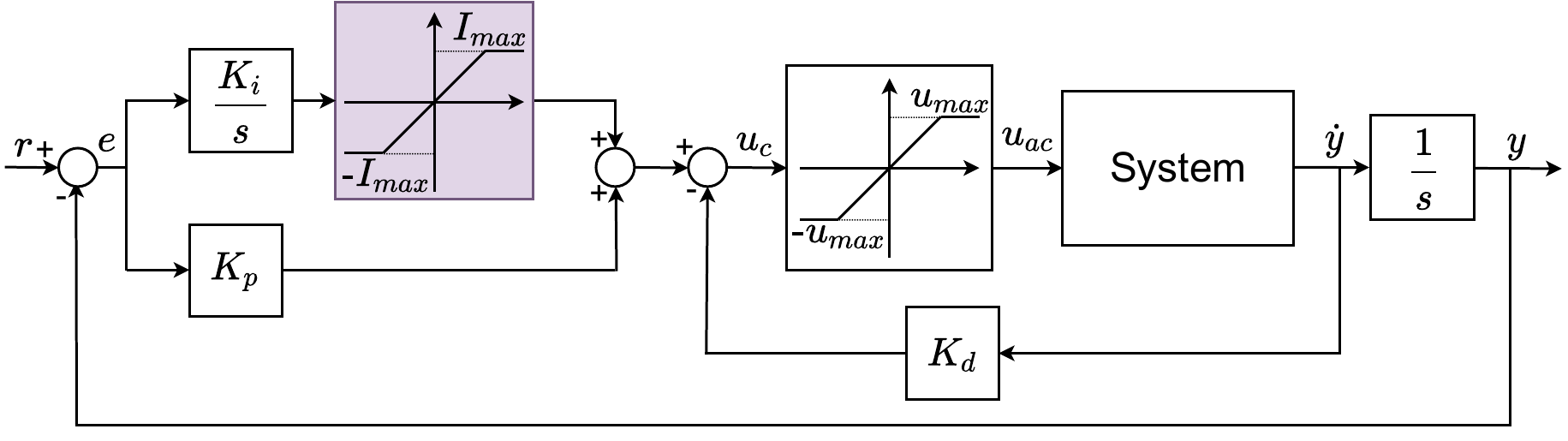}}
        \caption{A PID controller modified with integral clipping.}
	\label{fig:Integral_Clipping_block_diagram}
\end{figure}

Early methods for handling integrator windup aimed to limit or control the integrator’s action within the control loop. One of the simplest methods applies saturation to the output of the integrator branch in PI or PID controllers. This modification, schematically illustrated for a cascade PID controller with rate feedback in Figure \ref{fig:Integral_Clipping_block_diagram}, has long been used in autonomous vehicles and process control, and is commonly referred to as \textit{integrator clipping} or \textit{integral limiting}. In these methods, a limiter is applied to the integrator output, and its threshold is typically set as a proportion of the overall control signal boundaries. For instance, $50\%$-$80\%$ of the actuator capacity may be allocated to the integrator control signal.
There are other solutions, such as freezing the integrator by setting it to zero once saturation occurs, known as \textit{conditional integrating} and \textit{integrator clamping}. More advanced logics to freeze the integrator can be found in \cite{penghanus1996simpleAW,hodel2001AWclassicapproaches,visioli2003AWclassicapproaches,Visioli2006AWclassicapproaches}.

\subsubsection{Anti-reset windup or integrator resetting:}
Another well-known and practical approach to deal with saturation is the family of anti-windup methods. AW techniques typically operate based on the difference between the computed control signal $u_c$ and the actuated control signal $u_{ac}$, and are only active when saturation occurs \cite{tarbouriech2009ModernAWreview1}. Among these algorithms, a solution known as \textit{anti-reset windup} or \textit{back-calculation}, and sometimes referred to as \textit{integrator resetting}, has become particularly attractive. 
\begin{figure}[t!]
	\centerline{\includegraphics[width=0.5\textwidth]{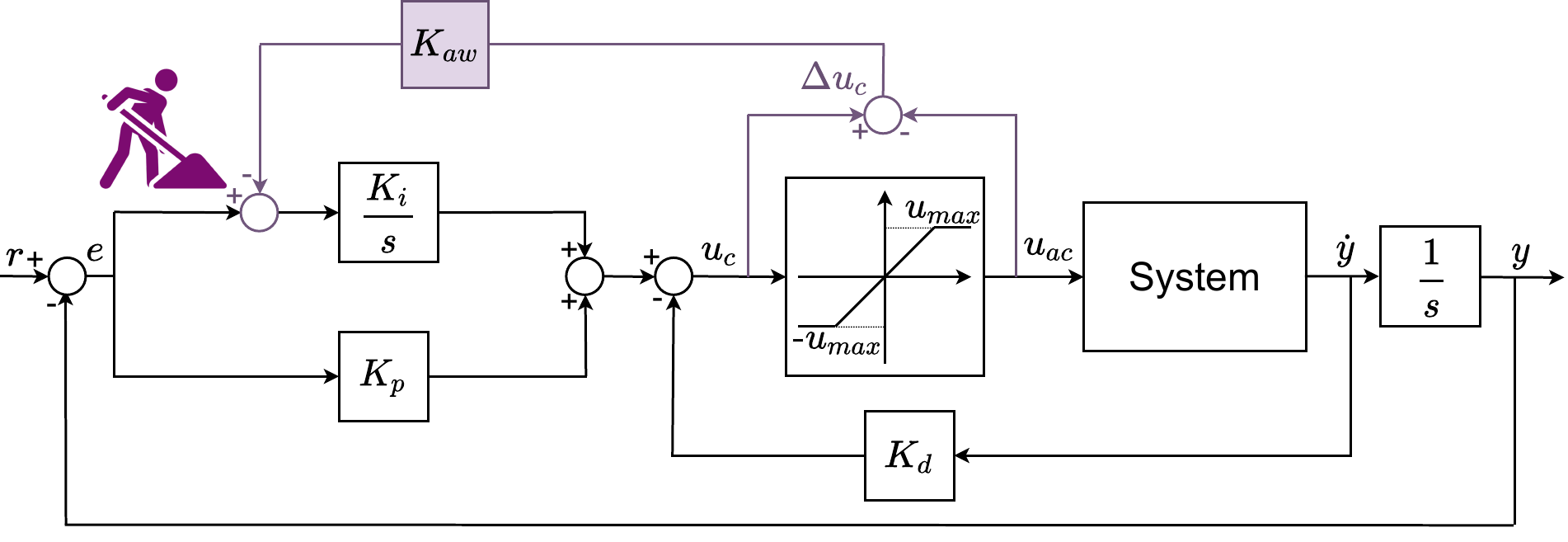}}
        \caption{A PID controller modified with integral resetting.}
	\label{fig:Integral_resetting}
\end{figure}
The concept of this approach, as applied to the cascade PID controller, is presented in Figure~\ref{fig:Integral_resetting}. As can be seen from the figure, the idea is to reduce the input to the integrator (typically the error signal, $e$) in proportion to the discrepancy between the two sides of the actuator, i.e., $\Delta u_c = u_c - u_{ac}$, commonly referred to as the \textit{control deficiency}. In other words, the back-calculation mechanism, through a feedback gain $K_{aw}$, feeds back this control deficiency signal to shave off the excess integration as illustrated in Figure~\ref{fig:Integral_resetting}. 
\subsection{Modern dynamic anti-windup controllers:}
\begin{figure}[t!]
	\centerline{\includegraphics[width=0.5\textwidth]{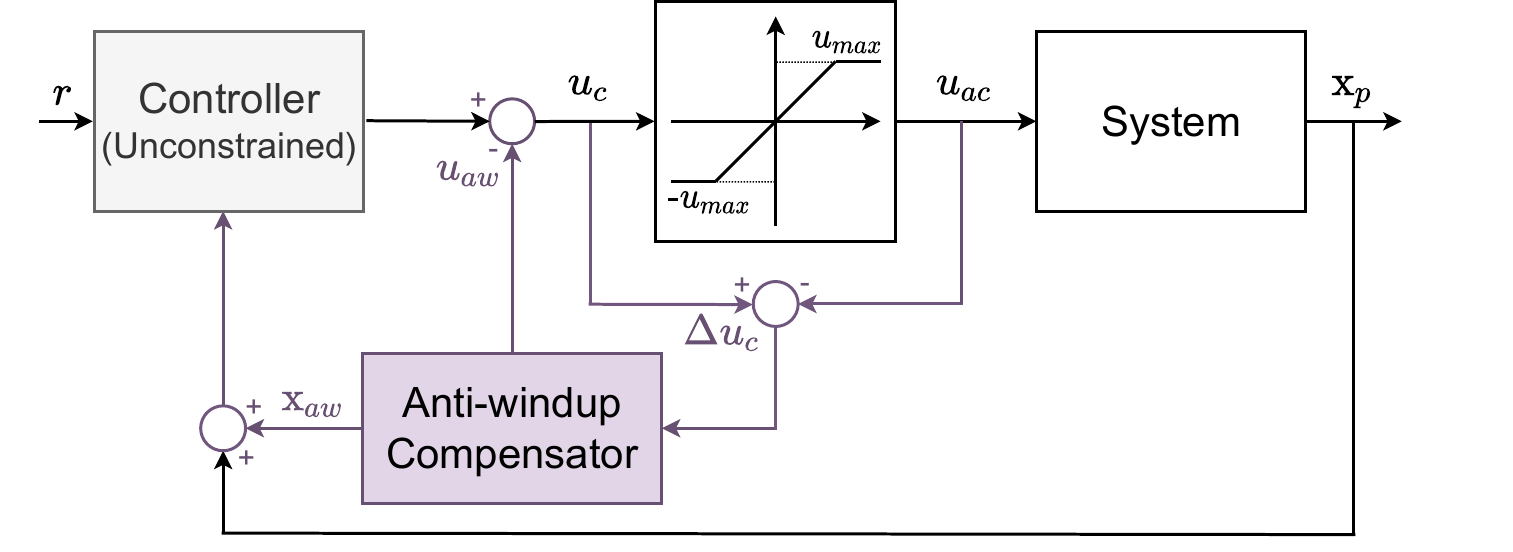}}
        \caption{Common structure of modern AW compensators.}
	\label{fig:Modern_AW_schematic}
\end{figure}

\vspace{-3pt}
\label{section:MAW_amplitude_saturation}
Another category of AW approaches is comparatively new (developed mainly in the 21st century) and is termed \textit{Modern} Anti-Windup (MAW) \cite{galeani2009ModenAWtutorialreview2}. Early efforts in the 1990s and 2000s aimed to formalise anti-windup design using robust and nonlinear control theory. See \cite{kapasouris1988AW,kothare1994AW,weston2000AW} and other references therein, which laid the foundation for the development of MAW methods in recent years. Some catastrophic incidents, in which underestimation of actuator (rate) saturation was identified as a root cause \cite{brieger2009flightATTAS}, along with discussions on practical control requirements in the community, catalysed this process \cite{Stein2003Respect}. The historical development of MAWs is reviewed in detail in \cite{tarbouriech2009ModernAWreview1,galeani2009ModenAWtutorialreview2} and we only introduce their common structure as illustrated in Figure \ref{fig:Modern_AW_schematic}.
MAWs are often dynamic compensators rather than simple static correction terms and can be represented as:
\begin{align}
\dot{x}_{aw}(t) &= (A_p + B_p F_{aw}) x_{aw}(t) + B_p \, \Delta u_c(t), \label{eq:dynamic_AW_eq1} \\
u_{aw}(t) &= F_{aw} \, x_{aw}(t), \\
y_{aw}(t) &= C_p \, x_{aw}(t),
\end{align}
where $x_{aw}$ symbolises the states of the anti-windup compensator. These compensators introduce a two-way modification to the system: $u_{aw}$ adjusts the control signal, while $y_{aw}$ or $x_{aw}$ alter the system states entering any baseline unconstrained controller. The matrices $A_p$, $B_p$, and $C_p$ are the standard system matrices defined in Eq.~\ref{eq:state_space_1}. These compensators have shown success in practice \cite{brieger2009flightATTAS,sofrony2010AW2rate}, and the problem is transformed to synthesising a model-based compensator to find $F_{aw}$ \cite{tarbouriech2009ModernAWreview1,galeani2009ModenAWtutorialreview2}. MAWs utilise robust and optimal control theories in their design, offering formal stability proofs and applicability to multiple-input multiple-output (MIMO) systems. Various techniques have been proposed to handle both A\&RSat in this context, and research in this area continues \cite{zaccarian2011AWBook1,tarbouriech2011AWBook2,lorenzetti2022ratesaturation}. However, due to the relative complexity of their design process, they are not widely adopted within the engineering community, where the tendency has shifted towards MPC or other online optimisation-based techniques, which may also present practical challenges. Hence, engineers may seek simpler controller modifications for common systems that can operate effectively with minimal alterations.
\subsection{Rate Saturation Problem in Control Systems}
The problem of \textit{rate saturation}, despite its critical importance, has received relatively little attention in the literature, particularly within classical AW approaches. For instance, a 1961 public report by the NASA Flight Research Centre made the following observation regarding instability in the lateral controller of an F-104A aircraft \cite{weil1961ratesaturation}:

\vspace{3pt}
\textit{"It was found that the use of a relatively high yaw-damper gain produced \textbf{both the deflection and rate saturation} which rendered the yaw damper completely ineffective in the initial NASA flight experience. Later, flights with reduced damper gain and a 100\% increase in the available yaw-damper rate produced a marked decrease in the motions obtained following intentional throttle retardation."}
\vspace{6pt}

This shows that the problem was already recognised at the time, with high-gain (fast) control identified as a key factor increasing the likelihood of saturation and instability. Consequently, reducing controller gains or modifying/replacing the hardware was proposed as an effective solution. These techniques dominated early strategies for handling saturation, typically slowing system performance through gain reduction or applying smoother reference signals to avoid triggering saturation. In another study conducted in 1980, Hanus \cite{hanus1980preventing} briefly discussed the windup problem for rate-saturating actuators. Subsequently, CAW solutions for rate-saturated actuators were addressed sparingly in \cite{penghanus1996simpleAW,hippe2006windupratesaturationbook,flesch2017unifiedratesatnormeyrico}, with \cite{flesch2017unifiedratesatnormeyrico} providing a more recent contribution, after which research on CAW for A\&RSat appears to have declined. The work in \cite{flesch2017unifiedratesatnormeyrico} proposed simple yet effective solutions for A\&RSat, although further analysis of rate saturation effects on closed-loop stability remained an open topic. Nevertheless, a turning point in control theory occurred during the 1990s, when formal and optimisation-based design methods became more prominent in the literature \cite{kapasouris1988AW,duda1997PIO,stoorvogel1999ratesaturation,galeani2006ratesaturation,lorenzetti2022ratesaturation} (see also \cite{tarbouriech2009ModernAWreview1,galeani2009ModenAWtutorialreview2} for further references). Consequently, MAW techniques have provided strong formal solutions at the cost of increased complexity, whilst simple CAW methods handling both amplitude and rate saturation have received comparatively limited attention in recent years. This gap was the core motivation to highlight the importance of CAW and their feasibility for a wide range of applications, particularly in industries with limited computational, time, modelling, or expertise resources required to implement advanced methods.

\section{The Proposed Anti-windup Solutions}
\vspace{-3pt}
\label{section:proposed_techniques}
We propose two techniques to modify PID and LQI controllers to enhance their performance in dealing with rate saturation. The strength of these methods lies in the fact that they only need the actuator input-output signals, which are often available in modern actuators, eliminating the need for internal models in implementation. They are simple to implement and can operate on versatile systems with appropriate tuning.

\subsection{Modified PD Controller with AW (PD\_AW)}
\label{subsection:PID_AW}
\begin{figure*}[t!]
	\centerline{\includegraphics[width=0.9\textwidth]{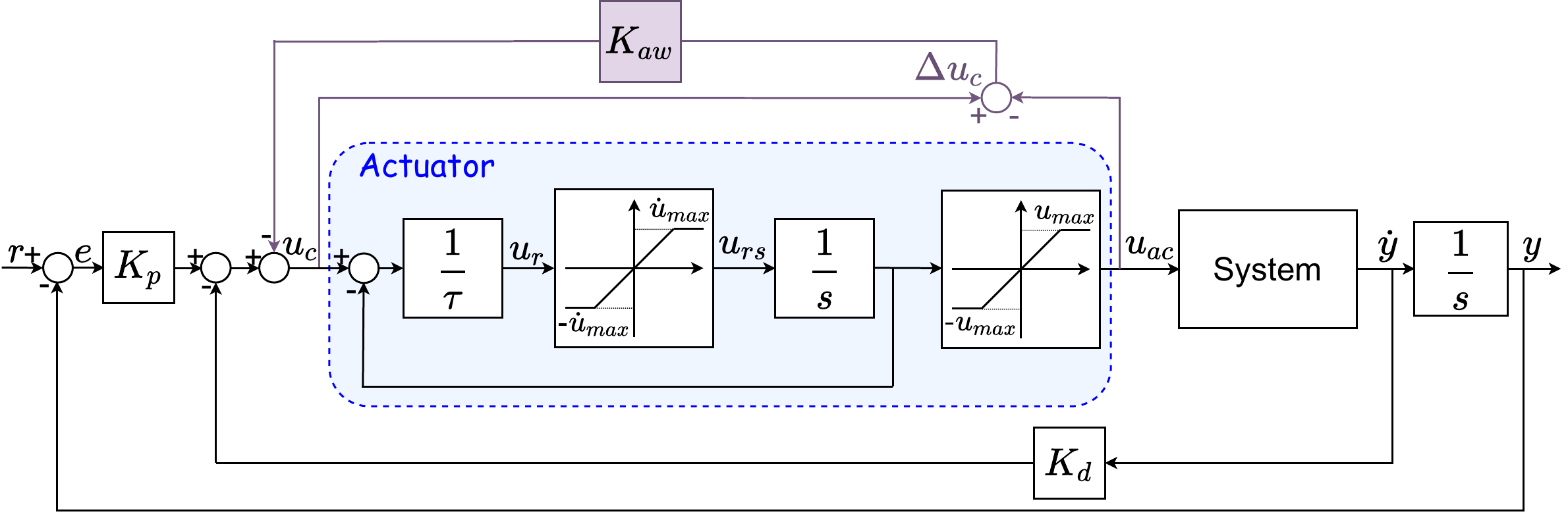}}
        \caption{Block-diagram of PD\_AW controller (a PID version requires clipping the integrator output).}
	\label{fig:PID_RS_block_Diagram}
\end{figure*}

\vspace{-3pt}
The first method augments the AW capability for handling A\&RSat in the cascade PD controller (referred to as PD\_AW), and its block diagram is shown in Figure~\ref{fig:PID_RS_block_Diagram}. The concept is similar to the back-calculation approach; however, it uses the actuator input-output signals (including both R\&Sat effects) and modifies the \textit{controller output} through an AW feedback gain, $K_{aw}$. It should be noted that if a PID controller is used, the output of the integrator branch should also be clipped, as discussed in Section~\ref{section:ad-hoc-amplitude-AW} and shown in Figure~\ref{fig:Integral_Clipping_block_diagram}. A PID controller with integrator-resetting cannot cope with this problem. The motivation for modifying the PD controller is that windup can occur even without an integrator term, while nominal tracking in the absence of saturation can still be achieved using only PD action. Hence, the PD\_AW control signal is given by:
\begin{equation}
    u_c(t) = K_p e(t) - K_d \dot y(t) - K_{aw} \Delta u_c
    \label{eq:PD_control_signal}
\end{equation}
Its implementation is presented in the next section.
%
\subsection{Modified LQI Controller with AW (LQI\_AW)}
\label{subsection:LQI_AW}
\begin{figure*}[t!]
\centerline{\includegraphics[width=0.9\textwidth]{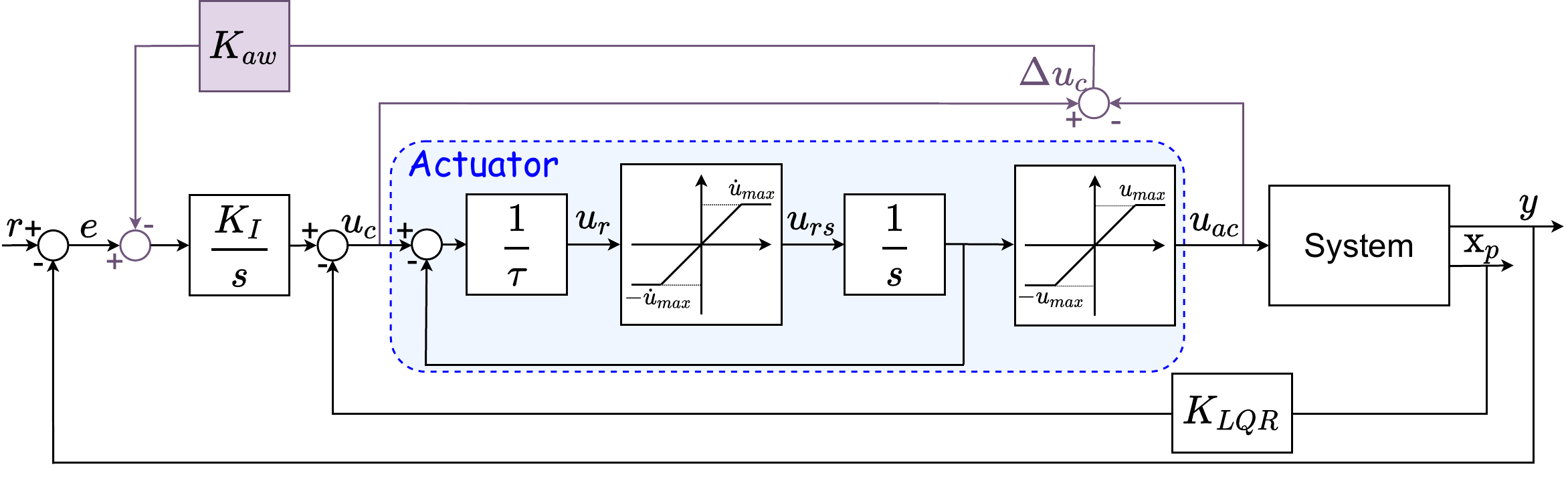}}
        \caption{Block-diagram of LQI\_AW.}
	\label{fig:LQI_RS_block_diagram}
\end{figure*}

\vspace{-3pt}
The second approach, LQI\_AW, follows the integrator resetting concept shown in Figure~\ref{fig:Integral_resetting}, with the main difference that it back-calculates and feeds back the full actuator input–output signal to the integrator. The block diagram of this method is illustrated in Figure~\ref{fig:LQI_RS_block_diagram}.  
LQI is a popular controller in autonomous vehicle applications due to its PID-like structure, state feedback approach, and optimal tuning capability. To design an LQI, an augmented system of the plant in Eq.~(\ref{eq:state_space_1}) is formed as:
\begin{align}
\underbrace{\begin{bmatrix}\dot{e}_I \\ \dot{x}_p\end{bmatrix}}_{\dot{x}}
&=
\underbrace{\begin{bmatrix}0 & -C_p \\ 0 & A_p\end{bmatrix}}_{A}
\underbrace{\begin{bmatrix}e_I \\ x_p\end{bmatrix}}_{x}
+
\underbrace{\begin{bmatrix}0 \\ B_p\end{bmatrix}}_{B} u
+
\underbrace{\begin{bmatrix}I \\ 0\end{bmatrix}}_{B_r} r, \\
y &= \underbrace{\begin{bmatrix}C_p & 0\end{bmatrix}}_{C} x.
\label{eq:augmented_system}
\end{align}
Here, $(A, B, C)$ are the augmented system matrices and $B_r$ is the reference input matrix. The additional state $e_I$, defined from the tracking error $e(t) = r(t) - y(t)$, is given by $e_I(t) = \int_0^t e(\tau)\,d\tau$.  
Applying the optimal linear–quadratic approach with the following cost function:
\begin{equation}
J = \int_0^\infty (x^\top Q x + u^\top R u)\,dt,
\end{equation}
where $Q \succeq 0$ and $R \succ 0$, can yield the following control law for the baseline controller:
\begin{equation}
u_c(t) = -K_x x(t),
\end{equation}
where $K_x$ combines feedback from both the plant states $x_p$ and the integral state $e_I$. The gain is decomposed as:
\begin{equation}
K_x = \begin{bmatrix} K_I & \vert & K_{x_p} \end{bmatrix},
\end{equation}
with $K_I$ acting on $e_I$ and $K_{x_p}$ on $x_p$. Thanks to its optimal tuning, clear structure, and suitability for MIMO systems, LQI is widely used in control applications \cite{lavretsky2024robust}. Therefore, a simple and effective extension of the LQI controller to handle A\&RSat is of significant interest. The modified LQI\_AW law can be written as:
\begin{equation}
u_c(t) = K_I\!\bigl(e(t) - K_{aw}\,\Delta u_c(t)\bigr) - K_{x_p} x_p(t),
\end{equation}
which differs from the PD\_AW control signal in Eq.~(\ref{eq:PD_control_signal}).
\section{Simulation Results}
\vspace{-3pt}
\label{section:simulation_results}
The system under study is the heading (yaw) controller of the REMUS Autonomous Underwater Vehicle (AUV) \cite{PresteroAUV}. A second-order model of the yaw dynamics at an operating speed of 1~m/s is expressed in state-space form as follows:
\begin{align}
\begin{bmatrix}
\dot{\psi} \\
\dot{r}
\end{bmatrix}
&=
\begin{bmatrix}
0 & 1 \\
0 & -2.16
\end{bmatrix}
\begin{bmatrix}
\psi \\
r
\end{bmatrix}
+
\begin{bmatrix}
0 \\
1.98
\end{bmatrix}
u_c, \\
y &= 
\begin{bmatrix}
1 & 0
\end{bmatrix}
\begin{bmatrix}
\psi \\
r
\end{bmatrix}.
\end{align}
Here, $x_p = [\psi \;\; r]^T$ represents the system states (yaw angle and yaw rate). The control input $u_c$ corresponds to the rudder deflection ($\delta_r$ in the nonlinear model). A deliberately challenging actuator model with a time constant of $\tau = 0.1~\text{s}$ and amplitude and rate constraints of $|u_c| \le 20~(\mathrm{deg})$ and $|\dot{u}_c| \le 30~(\mathrm{deg/s})$ is considered.

The system is controlled by three methods:  
i) \textbf{PD\_AW}, based on Section~\ref{subsection:PID_AW}, with $K_p=8$, $K_d=6$, and $K_{aw}=4$;  
ii) \textbf{LQI\_AW}, from Section~\ref{subsection:LQI_AW}, with weights $Q=\mathrm{diag}(1000,50,25)$, $R=1$, and $K_{aw}=4$;  This identical selection of the AW gains in both PID and LQI highlights the simplicity of the tuning process.
and iii) a \textbf{constrained MPC} controller optimising the following discrete-time cost function for the system discretised at sampling time $T_s=0.01$~s:
\begin{equation}
\begin{aligned}
J &= \sum_{k=0}^{N_y-1} \big( (r(k)-y(k))^2 + \lambda\, d u_c(k)^2 \big), \\
\text{s.t. } & |u_c(k)| \le u_\mathrm{max}, \ |\dot{u}_c(k)| \le \dot{u}_\mathrm{max}.
\end{aligned}
\end{equation}
where $k=0,\dots,N_y-1$ is the discrete-time index, $N_y$ is the output horizon ($\lambda=0.1$, $N_y=120$), and $du_c(k) = u_c(k)-u_c(k-1)$ is the control increment.
 All controllers are tuned to achieve their best smooth performance under both constrained and unconstrained conditions. 

\begin{figure*}[t!]
	\centerline{\includegraphics[width=0.98\textwidth]{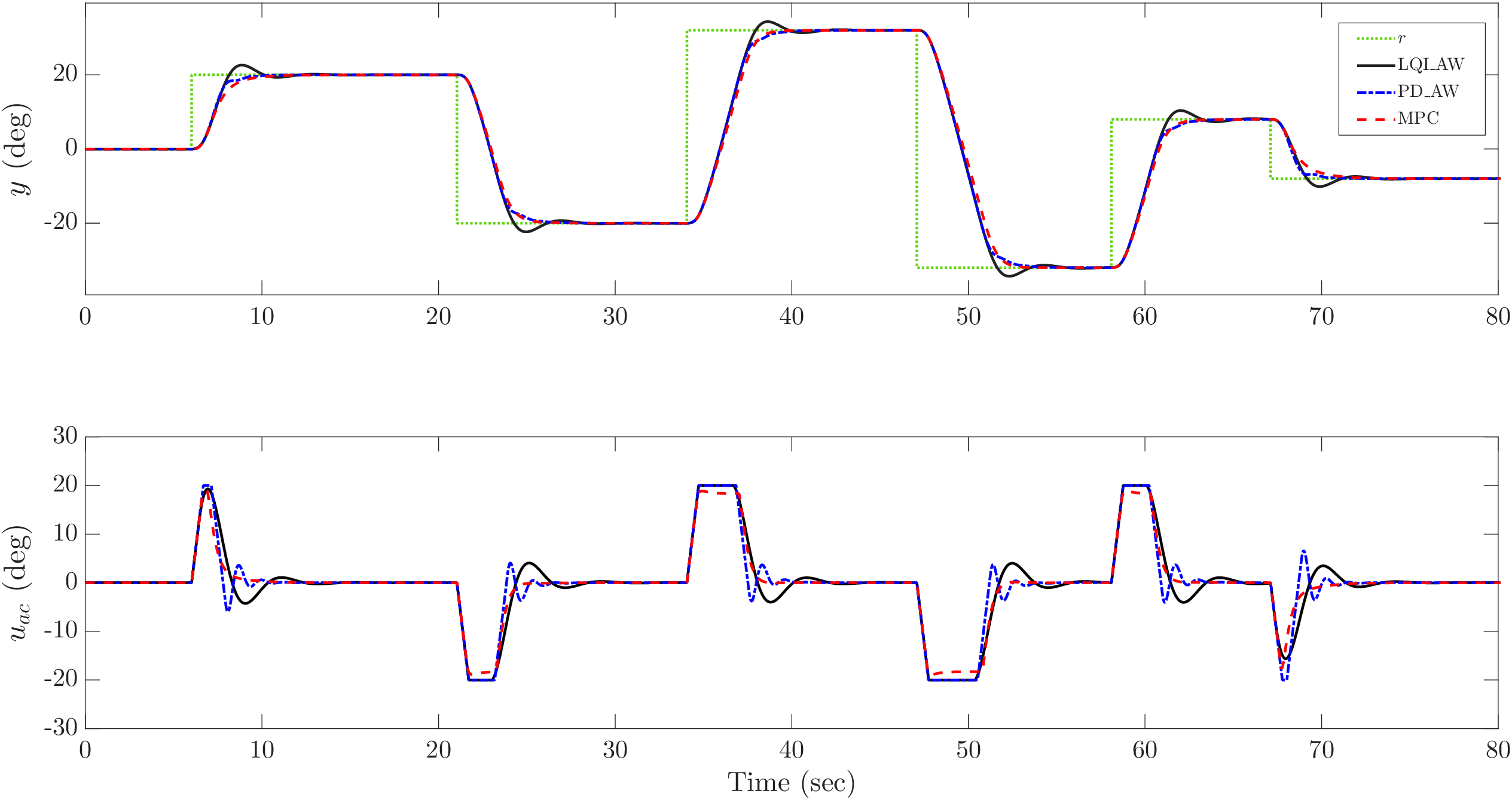}}
        \caption{Tracking and control signal comparison for PD\_AW, LQI\_AW, and MPC under demanding stepwise setpoints.}
	\label{fig:Rate_saturation_AW_comparison}
\end{figure*}

The results of the three controllers for the AUV heading dynamics under a demanding stepwise setpoint test scenario are compared in Figure~\ref{fig:Rate_saturation_AW_comparison}. It must be noted that all non-constrained versions of the controllers become unstable in this condition. These controllers maintain their tracking ability, while their transient responses remain smooth and acceptable, and from the tracking plot it is evident that the performance of all approaches is very similar. The bottom plot, showing the corresponding control signals, also indicates similar trends with minor discrepancies in transient conditions. The effect of rate saturation is visible in the control signal, resulting in slower transients for larger setpoint changes. Overall, based on Figure~\ref{fig:Rate_saturation_AW_comparison}, all three controllers exhibit comparable performance. 

For any applied controller to be successful in practice, three categories of metrics must be satisfied: 1) tracking performance, 2) control activity, and 3) robustness  (\cite{sarhadi2025standard}). Hence, for further analysis, five applicable metrics ($ISE$, $IACE$, $IACER$, $GM$, and $DM$) are employed. While other measures could be considered, metrics such as $ITAE$ are not suitable for varying stepwise setpoints, which is why the chosen metrics are preferred. Moreover, all cumulative metrics are normalised by the simulation time $T_f = 80$ to create smaller and more interpretable values.
Metrics for all controllers are presented in Table~\ref{table:AW_comparison}, which also supports the findings illustrated in Figure~\ref{fig:Rate_saturation_AW_comparison}. The tracking performance, as measured by $ISE$, is similar across all controllers, with PD\_AW and LQI\_AW performing slightly better. Conversely, MPC requires less control effort, as indicated by both $IACE$ and $IACER$ (more significant in $IACER$). Both PD\_AW and LQI\_AW exhibit higher gain margins, with LQI\_AW being superior, while MPC's robustness remains acceptable. MPC tolerates a significantly higher delay uncertainty, whereas the delay margins of the other two controllers are adequate given the system characteristics. Similar results are observed across different test scenarios, demonstrating that these simple approaches deliver good performance with lower complexity. Simulation files are shared openly for further analysis and comparison\footnote{\textcolor{blue}{\url{https://github.com/Psarhadi/Rsat-Classic-AW}}}\!.

These results highlight that the simple yet effective AW modifications applied to the PD and LQI controllers, each requiring only one additional gain, are capable of handling both amplitude and rate saturation. Due to the marginal stability of the open-loop system, the requirements of typical MAW approaches \cite{sofrony2010AW2rate} cannot be met, and thus these methods may not be directly applicable to this system. We acknowledge that various advanced MAW techniques exist to address such issues, and our intention is not to disregard them. However, given the comparable performance to even the MPC controller, one may opt for the simpler PD\_AW or LQI\_AW structures. This choice nonetheless warrants further analysis, as the findings may not generalise to all systems.

\begin{table}[t!]
\caption{Quantitative comparison of controllers}
\label{table:AW_comparison}
\centering
\resizebox{0.465\textwidth}{!}{
\begin{NiceTabular}{lccc}
\toprule
\rowcolor{MB} 
\textbf{Criteria} & \textbf{PD\_AW} & \textbf{LQI\_AW} & \textbf{MPC} \\
\midrule
\rowcolor[HTML]{FFFFFF} 
$ISE = \int_0^{T_f} e^2(t)\,dt / T_f$ & 21885.7 & 21933.1 & 22762.3 \\
\rowcolor[HTML]{E0F0FF} 
$IACE = \int_0^{T_f} |u_{ac}(t)|\,dt / T_f$ & 342.69 & 393.93 & 316.69 \\
\rowcolor[HTML]{FFFFFF} 
$IACER = \int_0^{T_f} |\dot{u}_{ac}(t)|\,dt / T_f$ & 445.35 & 366.18 & 280.38 \\
\rowcolor[HTML]{E0F0FF} 
Gain Margin - $GM (s)$ & 6.5 & \textgreater 10.5 & 5.5 \\
\rowcolor[HTML]{FFFFFF} 
Delay Margin - $DM (s)$ & 0.12 & 0.37 & 1.45 \\
\bottomrule
\end{NiceTabular}
}
\end{table}

\section{Conclusion}
\label{section:conclusion}
\vspace{-3pt}
This paper revisited the problem of amplitude and rate saturation from a classical anti-windup design perspective. Simple modifications to two popular controllers, PID and LQI, were proposed to enable them to handle these challenging nonlinearities. This topic has been relatively inactive in recent years, particularly regarding the treatment of rate saturation. Simulation results on AUV dynamics demonstrated that these solutions achieve suitable performance, comparable with MPC. The paper's intention was not to mandate the use of any particular approach, but to address the needs of the engineering community, which often seeks simple yet effective solutions that can modify a controller with a few (here, one) tuning knobs. Hence, the choice of controller can be guided by the designer's preference. We also encourage further theoretical research, including stability analysis and extension to MIMO systems, in line with the trend of applying such widely used simpler controllers in industry \cite{skogestad2023advanced}.

\bibliographystyle{IEEEtran}
\bibliography{cdcconf.bib}
\end{document}